# Welfare estimations from imagery: A test of domain experts' ability to rate poverty from visual inspection of satellite imagery

**Wahab**, **Ibrahim** & **Hall**, **Ola**

Department of human geography, Lund University, Sweden

Ibrahim.wahab@keg.lu.se, Ola.hall@keg.lu.se, twitter: Ohall @ohall20

## Abstract

*Background*: Rapidly and yet accurately estimating welfare levels at different spatial scales is critical to ensuring that no region is left behind in the quest for poverty reduction and eradication. Useful as the traditional workhorse of household surveys are, they are expensive to implement and often infrequent in time. Recent advances in remote sensing (mainly satellite imagery) and Artificial Intelligence (in the fields of machine learning, deep learning, and transfer learning) have led to increased accuracies in poverty and welfare estimation. These systems are, however, largely opaque in terms of explaining how these impressive results are achieved. To fulfil the need for explainable AI, domain knowledge; human experts with adequate contextual knowledge of poverty features becomes essential.

*Methods*: The present study uses domain experts to estimate welfare levels and indicators from high-resolution satellite imagery. We use the wealth quintiles from the 2015 Tanzania DHS dataset as ground truth data. We analyse the performance of the visual estimation of relative wealth at the cluster level and compare these with wealth rankings from the DHS survey of 2015 for that country using correlations, ordinal regressions and multinomial logistic regressions.

*Findings*: Of the 608 clusters, 115 (19%) received the same ratings from human experts and the independent DHS rankings. For 59% of the clusters, experts' ratings were slightly lower (Md = 2.50, n = 358) than DHS rankings (Md = 3.00, n = 135), z = -11.32, p = <0.001, with a moderate effect size, r = -0.32. On the one hand, significant positive predictors of wealth are the presence of modern roofs and wider roads. For instance, the log odds of receiving a rating in a higher quintile on the wealth rankings is 0.917 points higher on average for clusters with buildings with slate or tile roofing compared to those without. On the other hand, significant negative predictors included poor road coverage, low to medium greenery coverage, and low to medium building density. Other key predictors from the multinomial regression model include settlement structure and farm sizes.

*Significance*:  These findings are significant to the extent that these correlates of wealth and poverty are visually readable from satellite imagery and can be used to train machine learning models in *poverty predictions. Using these features for training will contribute to more transparent ML models and, consequently, explainable AI.*

## 1. Introduction

Reducing and eradicating poverty has been and continues to be an overarching ideal for the world. This is exemplified in how it is a common thread for many of the UN's sustainable development goals (SDGs), with the underlying principle being that no one or region is left behind. A fundamental prerequisite for poverty eradication, however, is accurate measurement and identification of regions, villages, and households plagued with poverty (Li et al., 2022). This, among other things, will help determine areas and regions that need development assistance the most. Different approaches tend to be used in different contexts to estimate welfare levels. In more developed regions, income-based approaches tend to be preferred (Achia et al., 2010). The income approach is the commonest and, more effectively implemented approach, to a large extent, because of the availability and accessibility of reliable income data. The same cannot be said for most developing countries where income reporting and data are either non-existent or incomplete and, thus, unreliable. The expenditure approach to welfare measurement is unreliable due mainly to the cash-heavy nature of transactions in developing countries' economies. The income and expenditure approach to poverty measurement is also an inaccurate measure of poverty and welfare in developing regions due to their susceptibility to seasonal variation (Shaukat et al., 2020). It is against this background that the asset-based approach has become mainstream in the assessment of the welfare status of households and neighbourhoods.

Poverty measurement that relies on asset holdings tends to be more accurate as variables relating to assets often correlate well with poverty, especially the structural kind (Li et al., 2022). The appeal of the asset-based welfare measurement is that not only does household welfare strongly correlate with the rise and fall in asset holdings (Brandolini et al., 2010), but also these holdings are relatively easy to observe and measure. The traditional means to measure asset-based welfare is census or household surveys. The two commonest surveys are the Living Standard Measurement Survey (LSMS) and the Demographic and Health Survey (DHS). Such surveys are useful as they are usually nationally representative. Though they collect data at the household level, their coverage is patchy and they are often expensive to implement and, as a consequence, are carried out only every few years. For increased efficiency in resource allocation, the identification of places in need requires more flexible, rapid, and more precise tools and approaches for poverty mapping (Wisner et al., 2014)

In the last decade, there has been increasing use of artificial intelligence (AI), particularly the deep learning (DL) subtype, in this area to meet this growing need. The increased application of these tools can largely be attributed to three main developments: (1) the proliferation of labelled big data from multiple sources, (2) significant breakthroughs in computing power, and (3) advances in cloud computing and storage. The application of AI and DL models in this domain of economic development studies is often done in conjunction with high-resolution remote sensing data or call detail records (CDR) and often as a complement to more traditional survey tools. In the last decade, seminal studies in this area include Jean et al. (2016) and Xie et al. (2016) which relied on such multimodal datasets. Preceding attempts in this area have relied on artificial lights at night (ALAN) which have been shown to be a good proxy for the level of economic activity (Ghosh et al., 2010; Keola et al., 2015). Reliance on ALAN has, however, seen limited applications in recent times due mainly to the lower accuracy of this approach especially in developing regions (Yeh et al., 2020). It is in light of this that better performance and accuracy are being achieved in studies that combine daytime and night-time satellite data (Hofer et al., 2020; Lee & Braithwaite, 2022; Tingzon et al., 2019).

There continue to be questions, however, relating to the inherently 'black-box' nature of the different models for not only poverty mapping and analysis at the local level but also in other high-stakes decisions (Rudin, 2019). The inherent risks relate to the opacity of such models. A recent review of studies relying on satellite imagery and machine learning (ML) in the poverty and welfare domain (Hall et al., 2022) underscores the importance of domain knowledge for the explainability of such models. Relegating such important decisions as which regions, villages, and neighbourhoods receive special development assistance, to largely inexplainable and opaque models is not ideal. In the area of welfare analysis, pertinent questions relate to what features and characteristics are indicative of poverty and thus correlate to welfare predictions using AI models. We hypothesize that achieving any level of explainability requires human-machine collaboration. How well can human experts with experiential and socio-cultural knowledge of the context estimate welfare in the landscape? Can human and artificial intelligence work together to improve poverty prediction accuracy? The present study seeks to estimate poverty and welfare levels from very high-resolution satellite data by human experts with domain knowledge of the context.

## 2. Theoretical foundations of spatial poverty estimation

At the fundamental level is the question of who qualifies as the poor or what is poverty. This conceptual question has plagued development practices and policymakers for several decades. The task of defining poverty assumed greater relevance in the 1960s in the aftermath of the attainment of political independence by many former colonies (Lee & Braithwaite, 2022). Suffice it to state that defining poverty or even the approach to doing this, is a strongly contested issue and the different positions are not always necessarily complementary. One of the well-known theorizations in this endeavour is the capability approach by Sen (1982, 2000) who posits that the specification of a certain consumption norm or of a poverty line does only a part of the job. This notwithstanding, the World Bank still annually publishes its report based on a hypothetical dollar per person per day metric – currently, 1.9 USD per person per day – as the so-called poverty line while the United Nations produces its annual reports on the human development index – a composite of life expectancy, education and standard of living – to measure welfare levels in individual countries. There has been a clear move towards indices as the multidimensional nature of poverty has become more mainstream (Alkire & Foster, 2011).

A key development in this theoretical evolution has been the discovery of the spatial qualities of poverty and deprivation. The theory of spatial poverty can be traced back to the theory of environmental determinism – the idea that the physical environment predisposes regions and countries to particular development trajectories. While environmental determinism has largely been rejected in favour of environmental possibilism, the spatiality of poverty continues to grow in relevance (Bird et al., 2010; Burke & Jayne, 2008; Elwood et al., 2016). Like its antecedent, the spatial poverty theory holds that geographical location plays an important role in the formation and persistence of poverty (Zhou & Liu, 2022). It has, in turn, spawned other concepts such as spatial poverty traps which are empirically verifiable. In this regard, Liu et al. (2017) recently demonstrated the tendency of the poor to spatially agglomerate through the so-called 'island effect' by showing that poverty is often concentrated in areas with limited natural resource endowment, fragile ecosystems and other poor geographic conditions. Therein lies the intricate link between poverty, space and place (Milbourne, 2010).

In more recent times, the spatial resolution of poverty mapping has been increasing – from national and regional levels to village, neighbourhoods and even individual levels. It is at these finer levels that the analysis is most relevant for poverty alleviation efforts. Spatial identification of poverty which entails identifying poor areas and impoverished populations can reveal the heterogeneous and geographical character of poverty (Erenstein et al., 2010). Poverty may be viewed as absolute or relative, chronic (persistent) or transient, regional (place-based) or individual (people-based), and rural or urban with these binaries often related in one way or the other (Zhou & Liu, 2022). Yet, the various binaries of poverty manifest in different ways at the micro level and so measuring spatial poverty is a complex endeavour.

In developing countries, urban poverty has not garnered as much research interest as rural poverty, even though the urban poor continues to grow steadily. In the urban setting, poverty tends to exhibit certain principal characteristics: (1) Inadequate income [to afford necessities such as food, safe and sufficient water, (2) Inadequate, unstable or risky asset base [both in terms of education and housing for individuals, households and communities], (3) Inadequate shelter [poor quality, overcrowded, and insecure], (4) Inadequate provision of public infrastructure [piped water, sanitation, drainage, roads, footpaths etc], (5) Inadequate provision of basic services [schools, public transport, health clinics, communications services], among others (Satterthwaite, 2001). These often culminate in *socio-spatial fragmentation and segregation* in urban centres on the basis of the concept of the neighbour effect (van Ham et al., 2012). This manifests through a process whereby, the rich self-isolate to appropriate spatial privileges while the poor self-segregate to, among other things, escape the high cost of living (Liu et al., 2017; Otero et al., 2021). The confluence of the commodification of housing through increasingly high costs of land, coupled with the speculative interests of private developers and the marketization of social housing results in this phenomenon in the Global South (Rath, 2022). The unequal spatial distribution of physical and social infrastructure, fragmentation, and social exclusion through the neighbourhood effect – the idea that where an individual lives has significant effects on their life chances, over and above their individual characteristics – then results (Otero et al., 2021; van Ham et al., 2012). The net effect is the proliferation of poverty hot spots such as slums and ghettos which further perpetuate a vicious cycle of poverty.

In the area of welfare analysis, some of these characteristics and manifestations of poverty can be estimated from the landscape. It is through 'reading the landscape' that we can estimate general and relative poverty levels and also understand the social relations and structures which are not self-evident for approaches that rely on surveys or interviews (Widgren, 2006). It is pertinent to note, however, that certain 'soft' measures of poverty such as consumption may not be directly apparent even on very high-resolution satellite imagery compared to 'hard' measures such as farm sizes, housing and roofing type, which are often self-evident and tend to manifest through the neighbourhood effect. This is largely because increased incomes of households are often invested in mechanised farming, improved housing, and other consumer goods (Östberg et al., 2018). An adequate domain knowledge (Hall et al., 2022) and an understanding of the socio-cultural context are critical to an accurate reading of landscapes though. For example, while the quality of roofing material and status of housing roofs can be a strong indicator of the welfare status of a household in many regions in Sub-Saharan Africa (SSA), it can be quite misleading to work proceed on this assumption in some Asian countries where merely completing one's roof means one will have a higher tax burden. This notwithstanding, assets whether at the individual or community level, have been shown to be a reliable measure of welfare levels.

## 3. Data and methods
### 3.1 The Tanzania 2015 DHS dataset
The ground truth survey data is derived from the 2015/2016 Tanzania Demographic and Health Survey (TDHS) dataset which is the sixth and latest in the DHS series in the country. While the overall DHS programme is concerned with collecting and monitoring data on population, health and nutrition, it also collects data on households' living standards. This paper uses data on the welfare status of households surveyed in this campaign. The 6$^{th}$ TDHS uses a nationally-representative sample of 12,563 households, grouped into 608 clusters across the 30 regions of the country. For the present paper, the unit of analysis is the cluster which ranged in sample size from 12 to 22 with a mean of 21 (SD=1.55) households per cluster.

### 3.1.1 Indicator variables and the wealth index

The DHS Wealth Index is an asset-based index designed to compare relative, rather than absolute, economic wellbeing of households independently from education and health. The index is based on households' *ownership of assets* (such as radio, televisions, telephones, refrigerators, computers, bicycles, motorcycles, cars, bank accounts, house and lands); *access to services* (such as electricity, toilet facilities, and water sources, and sources of energy for cooking); and *building materials* (primarily flooring, walls and roofing materials). Responses to questions on assets, services, and enumerators' observations of construction materials in the DHS survey are considered more reliable than self-reported income- and expenditure-based estimation of welfare and economic wellbeing (Staveteig & Mallick, 2014). The combined wealth index of clusters, the dependent variable, has been grouped into quintiles: 'poorest', 'poorer', 'middle' 'richer' and 'richest'. While others, such as Shaukat et al. (2020) and Li et al. (2022), who have used this dataset have regrouped these quintiles into poor and non-poor, we opted to maintain the original categorizations to bring further nuance to the relative poverty ratings. Additionally, we also computed the median quintile score for each cluster which we hypothesize to represent the overall cluster score in terms of the level of relative poverty.

### 3.1.2 Geolocation of clusters

While it is now standard practice for DHS datasets to come with location data of clusters, the DHS programme anonymises it by adding noise to the location variable for ethical reasons. In the current dataset, urban clusters are displaced by a distance of up to two kilometres and up to five kilometres for rural clusters, with a further randomly selected one per cent of the rural clusters displaced by up to 10 kilometres (Burgert & Prosnitz, 2014). This introduces uncertainties in the predictions as we matched coordinates with clusters. We overcome this by 'correcting' the GPS data to the most likely location in an iterative manner on Google Maps. We achieve this by ensuring that we follow the guidelines for adding noise to the locations. That is, for rural locations, we assume that the nearest settlement within 5 km is the most likely 'actual' cluster and expand the search area to 10 km in the few cases where we could not find any possible settlement in the 5km radius. We then use this 'corrected coordinates' of clusters to extract cluster images.

### 3.2 Aerial image extraction and online survey

The Google Maps Platform hosts a set of APIs through which developers can retrieve data from the platform. It is notoriously difficult to find metadata about images provided in the service. Generally, images are from different sensors and combined into a mosaic of images taken over multiple periods. Sometimes seams between different images are possible to identify.

The corrected coordinates were fed into an R-script accessing the Google Maps platform and downloading corresponding ultra-high-resolution images. In total, 608 images were downloaded at zoom level 18. That corresponds to a pixel size of about 0.6 meters.

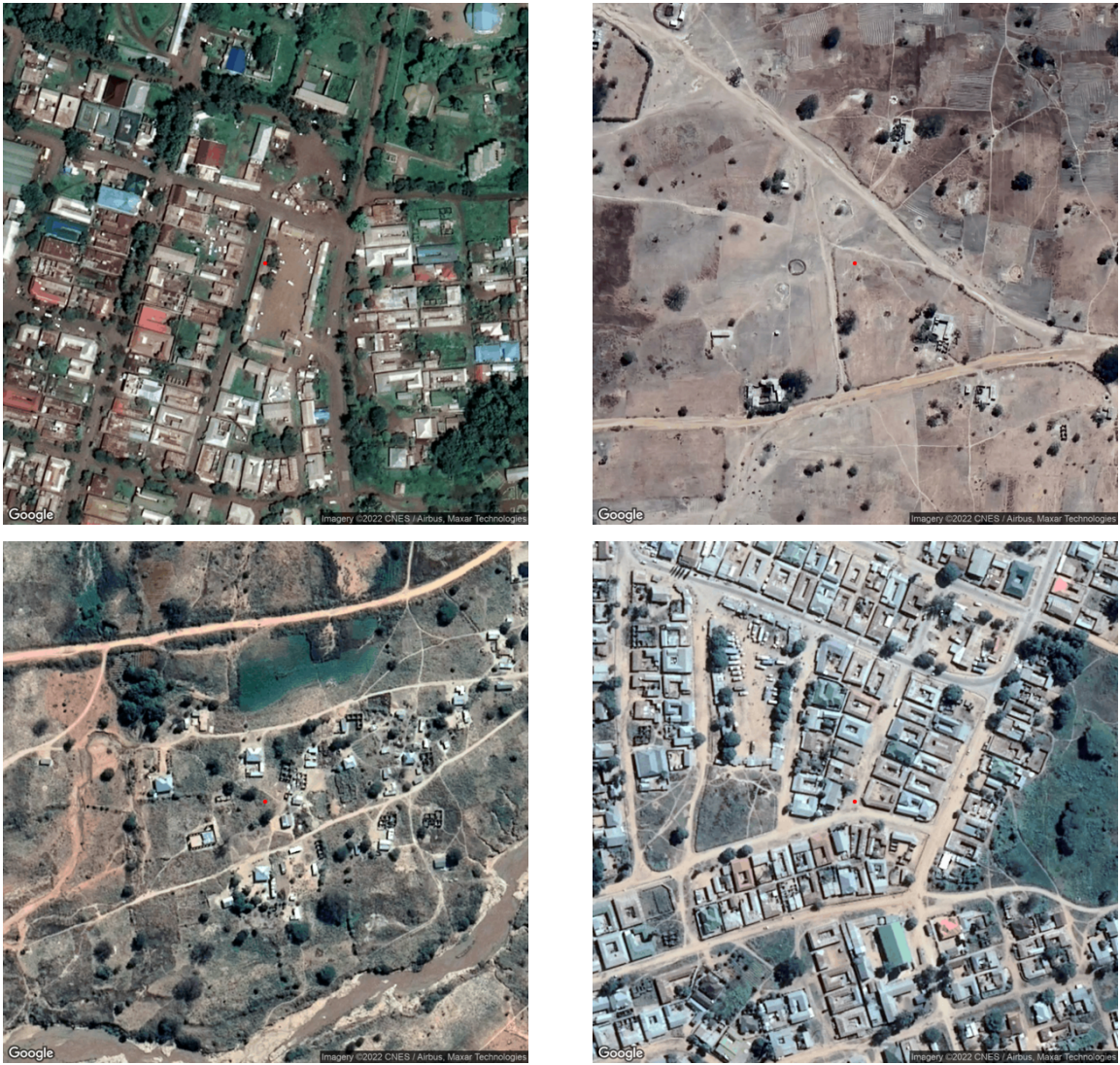

**Figure 1**. Example of images from four different locations in Tanzania. Note that images are reduced by 40% from original size.

The extracted images were then fed into the web-based Predict survey (https://predict.gis.lu.se/). A pilot for the survey run for two weeks using a group of master's students at Lund University to test the resilience of the platform as well as the validity and reliability of our experimental design before the main survey went live between March 1 and May 31, 2022. Our approach was to send personalised invitations to experts within our networks who have fieldwork experience in Africa in the area of development research, broadly defined. We had contemplated sending a general invitation to our networks or asking respondents to forward invitations to other experts they deem qualified but eventually decided against these two options. First, we supposed that a general, unsolicited email might not garner the same response rate as a personal one. Secondly, we wanted to control the sample of human experts who rate the images. Each expert is first presented with a batch of 30 images in a sequential manner and asked to rate each one as either 'poorer', 'poor', 'average', 'wealthy' or 'wealthier', to be comparable with the DHS quintiles. Raters were allowed to rate additional batches of 30 images if they so wished. Overall, the web portal received 2,174 ratings from 102 experts. The median rating for each cluster image was also taken as representing the relative poverty score for the cluster.

To shed light on the correlates of the ratings by experts, a criterion was developed, based on literature and discussions with experts with extensive knowledge of the Tanzanian context to classify and grade the images based on pertinent spatial features which could be discerned. On this, we rated the images on three broad criteria, namely, *housing features* (building types, roofing materials, roofing condition, and buildings' size); *landscape features* (settlement structure, building density, greenery coverage, dominant land use, and image colour scale) and *assets and infrastructure* (roads surface quality, roads width, roads coverage, vehicles presence, and farm sizes. Table 1 presents a detailed description of the sub-criteria we used for the classification. Each image was then manually inspected for these qualities and scored accordingly to find the correlates of the ratings by the domain experts.

Table 1. Literature and expert-based correlate features in images

| Broad criteria | Spatial features | Categories | Codes |
| --- | --- | --- | --- |
| **HOUSING FEATURES** | Size of buildings | single units | 1 |
| | | multiple units | 2 |
| | | unclassified | 99 |
| | Roofing materials | thatch roofing | 1 |
| | | aluminium roofing | 2 |
| | | slate roofing | 3 |
| | | unclassified | 99 |
| | Roofing condition | uncompleted | 1 |
| | | old | 2 |
| | | new | 3 |
| | Presence of tall buildings | no | 1 |
| | | yes | 2 |
| | | unclassified | 99 |
| **LANDSCAPE FEATURES** | Structure of settlements | clustered | 1 |
| | | scattered | 2 |
| | | gridded | 3 |
| | Density of built environment | low coverage (less than 25%) | 1 |
| | | medium coverage (between 25–74%) | 2 |
| | | high coverage (greater than 75%) | 3 |
| | Presence of trees/greenery | low (below 25%) | 1 |
| | | medium (25–75%) | 2 |
| | | high (above 75%) | 3 |
| | Dominant land use | bare land | 1 |
| | | agricultural | 2 |
| | | built up | 3 |
| | | industrial | 4 |
| | | commercial | 5 |
| | | unclassified | 99 |
| | Image colour | brownish | 1 |
| | | yellowish | 2 |
| | | greenish | 3 |
| **ASSETS AND INFRASTRUCTURE** | Presence of vehicles | no vehicles visible | 0 |
| | | few vehicles (1–2 visible) | 1 |
| | | many vehicles (more than 2 visible) | 2 |
| | Road surface quality | no roads | 0 |
| | | untarred roads | 1 |
| | | tarred roads | 2 |
| | Road width | small | 1 |
| | | medium | 2 |
| | | large | 3 |
| | | unclassified | 99 |
| | Farm sizes | small farms | 1 |
| | | Large farms | 2 |
| | | no farms | 3 |
| | Road coverage | low | 1 |
| | | medium | 2 |
| | | high | 3 |

All three datasets – the DHS survey data, experts' rating of relative poverty, and researchers' scoring of discernible features in images – were merged into one dataset using the cluster code as the link.

### 3.3 Analytical approach

Given that the ratings from the DHS dataset are expressed in quintiles, from poorest (1) to richest (5), and the same rating convention is followed in the domain experts' ratings, we consider the dependent variable, the wealth quintiles, as ordinal. We opt for the ordinal rather than interval scale because we cannot guarantee equal intervals between the quintiles, especially with the web-based ratings. That is, we cannot assure that the distance between poorest and poor is the same as that between rich and richest. We also tested for the normality of the distribution and find that while the ratings for urban clusters were normally distributed (skewness = -0.72, kurtosis = 0.04), those for rural clusters were not (skewness = 4.10, kurtosis = -0.30). We therefore employ the Spearman's rank correlation coefficient to test the relationship between the two set of ratings at the cluster level. The Spearman's rho is given as follows:

$$\rho = 1 - \frac{6\sum Di^2}{n[n^2 - 1]}$$

where ρ = Spearman's rho, D = the difference between the ranks for cluster I and n = number of observations.

Tied rankings were estimated using the Related Samples Wilcoxon Signed Rank Test, the preferred measure for definite scores which drops tied pairs (Scheff, 2016). An ordinal logistics regression was run to model the relationship between the welfare ratings and the image characteristics to determine the key features which influence wealth ratings. To do this, the variables for the image characteristics in Table 1 were converted and recoded with dummies as shown in Table 2.

Table 2: Recoded independent variables for the regression model

| Main indicator | Description of variable | Codes |
| --- | --- | --- |
| Building size | Multiple housing units | Present = 1, otherwise = 0 |
| Roofing material | Modern (slate) roofing | Dummy = 0 for thatch and aluminium roofing |
| Roofing condition | New roofs | Dummy for uncompleted or old roofs |
| Taller buildings | Presence of taller buildings | 1, otherwise (dummy) = 0 |
| Settlement structure | Spatially well-planned (gridded) settlements | 1, otherwise (dummy) = 0 |
| | Clustered settlement | 1, otherwise (dummy) = 0 |
| | Scattered settlements | 1, otherwise (dummy) = 0 |
| Density of built environment | Low building density (< 25% coverage) | 1, otherwise (dummy) = 0 |
| | Medium building density (25-75% coverage) | 1, otherwise (dummy) = 0 |
| | High building density (> 75% coverage) | 1, otherwise (dummy) = 0 |
| Greenery and trees presence | Low greenery coverage (< 25% coverage) | 1, otherwise (dummy) = 0 |
| | Medium greenery coverage (25-75% coverage) | 1, otherwise (dummy) = 0 |
| | High greenery coverage (> 75% coverage) | 1, otherwise (dummy) = 0 |

| Dominant land use | Bare lands | 1, otherwise (dummy) = 0 |
|---|---|---|
| | Agricultural fields | 1, otherwise (dummy) = 0 |
| | Built up surfaces | 1, otherwise (dummy) = 0 |
| Image quality | Greyish image discolouration | 1, otherwise (dummy) = 0 |
| Road surface quality | Tarred roads | 1, otherwise (dummy) = 0 |
| | Untarred roads | 1, otherwise (dummy) = 0 |
| Road sizes/width | Wider (medium or large) roads | 1, otherwise (dummy) = 0 |
| | Narrow (small) roads | 1, otherwise (dummy) = 0 |
| Road coverage | Low roads coverage | 1, otherwise (dummy) = 0 |
| | High roads coverage | 1, otherwise (dummy) = 0 |
| General farm sizes | Presence of farms | 1, otherwise (dummy) = 0 |
| | Presence of small farms | 1, otherwise (dummy) = 0 |
| | Presence of large farms | 1, otherwise (dummy) = 0 |
| Presence of vehicles | Presence of any vehicles (cars, tractors, etc) | 1, otherwise (dummy) = 0 |

## 4. Results

### 4.1 Summary description of human experts

As Table 3 shows, the sampled domain experts come with enormous experience. This shows in not only their educational attainment and age distribution – 43% are 45 or more years old and 50% hold a Ph.D. – but even more important, half of the respondents (N=51) have more than 10 years of fieldwork experience in SSA. In terms of the region of experience, 48% have Tanzania, Eastern or Southern Africa as their primary area of fieldwork experience.

Table 3. Summary statistics of web-based survey respondents

| **Variable** | **Categories** | **Distribution** |
|---|---|---|
| Age of respondents | *Up to 35 yrs old* | 31 |
| | *36 - 55 yrs old* | 53 |
| | *56 yrs old and above* | 18 |
| Gender | *Females* | 23 |
| | *Males* | 77 |
| | *Non-binary* | 2 |
| Educational qualifications | *Bachelors* | 17 |
| | *Masters* | 34 |
| | *Ph.D.* | 51 |
| | *1 - 5 yrs* | 24 |

| Length of experience (years) | 6 - 10 yrs | 27 |
|---|---|---|
| | 11 - 20 yrs | 31 |
| | Above 20 yrs | 20 |
| Region of experience | Tanzania | 14 |
| | Central and West Africa | 53 |
| | Southern and Eastern Africa | 35 |

Note: N=102 respondents.

Of the 608 clusters for the 2015 Tanzania DHS dataset, 428 were classified as rural with the remaining 180 classified as urban. By our iterative approach to geolocate the clusters, we estimate average dislocation of 3.09 (SD=2.66) km for rural areas and 1.15 (SD=0.77) for urban clusters. These fall within the DHS geographical displacement guidelines. Figure 2 (a) and (b) show the distribution of the wealth quintiles based on ratings by our domain experts and the DHS dataset, respectively. While comparable in terms of the quintile's categorisation, it must be noted that the unit of analysis for the DHS survey is the household (N=12,563) which then had to be aggregated to the cluster level to match the web-based survey which unit of analysis is the cluster level (N=608). As figure 2 a and 2b show, the distributions of both independent sets of ratings broadly follow an inverted U-shaped curve.

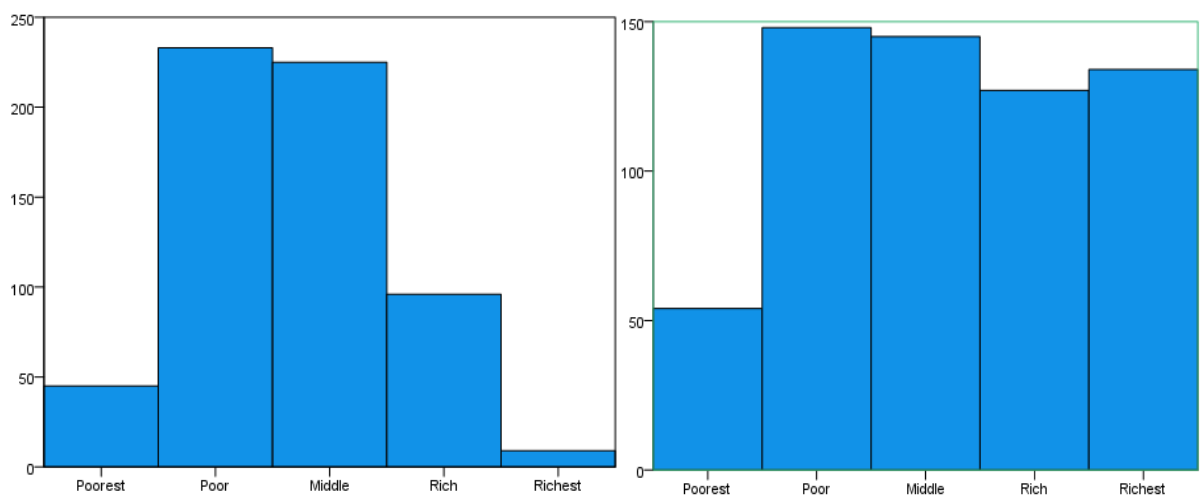

**Figure 2**: Wealth quintile distribution at the cluster level from (a) domain experts' ratings; (b) the 2015 Tanzania DHS at the household level (N=608 clusters).

### 4.2 Bivariate analyses of DHS quintiles and experts' scores of relative poverty

To analyse for the relationship between the poverty ratings from the DHS and those based on ocular estimation of relative poverty through the web-based survey at the cluster level, we conduct a Spearman's rank correlation test. Overall, the coefficient of determination of $r = 0.32$ for the wealth rankings from the web survey and DHS quintiles may appear weak based on conventional interpretations. However, given the nature of our dataset and the methodology we employed – spatially extrapolation and generalization that we have had to use due to the absence of data at the household level – we consider a statistically significant ($p = < 0.01$) coefficient of 0.32 as moderately strong. The relationship is a positive one which implies that there is a moderately strong direct relationship between median wealth ratings of clusters images and the wealth ranking quintiles from the DHS dataset at the cluster level.

Table 4: Bivariate relationship among web ratings and DHS wealth quintiles

| Variable | Expert web ratings | Mean DHS wealth quintiles (combined) | Mean DHS wealth quintiles (separated) | Median DHS wealth quintiles (combined) | Median DHS wealth quintiles (separated) |
|---|---|---|---|---|---|
| Expert web ratings | 1 | | | | |
| Mean DHS wealth quintiles (combined) | 0.313** | 1 | | | |
| Mean DHS wealth quintiles (separated) | 0.089* | 0.543** | 1 | | |
| Median DHS wealth quintiles (combined) | 0.315** | 0.973** | 0.540** | 1 | |
| Median DHS wealth quintiles (separated) | 0.107** | 0.571** | 0.933** | 0.579** | 1 |

Notes: N = 608, * = p<0.05, and ** = p<0.01 (2-tailed). *Combined* means that rural and urban clusters were undifferentiated, while *separated* implies rural and urban clusters were separated.

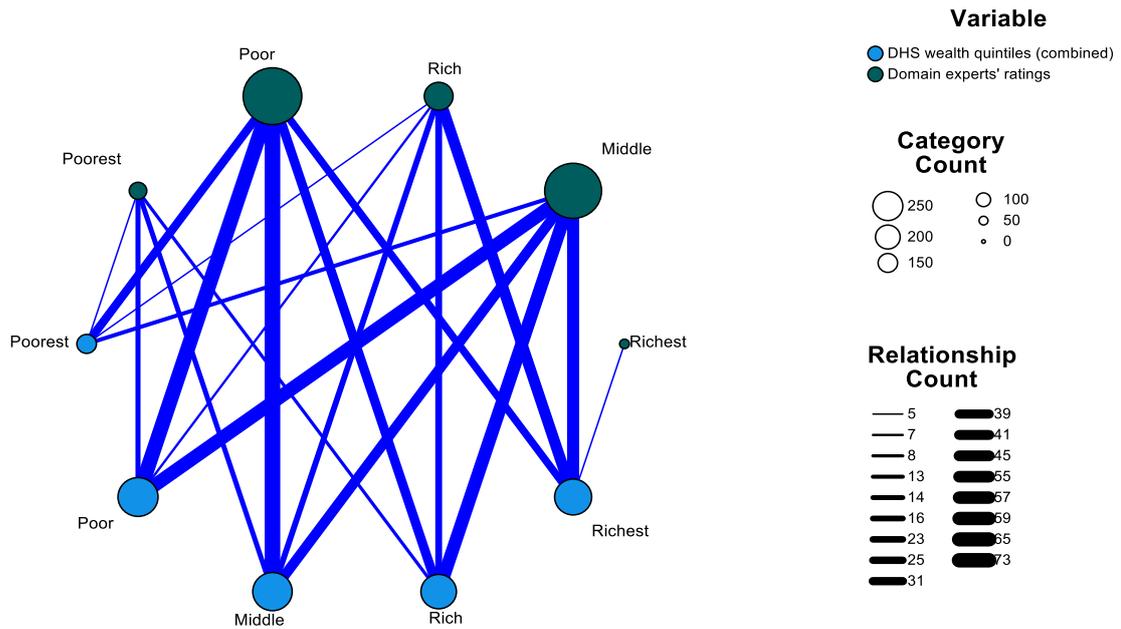

**Figure 3**: Relationship map of wealth rankings between DHS wealth quintiles and domain experts' ratings

Figure 3 shows the relationship between the different quintiles in the domain experts' ratings and those based on the DHS. On the one hand, all the clusters rated 'Richest' by domain experts were also only rated 'Richest' in the DHS wealth quintiles. Similarly, most of the 'Rich' clusters from the experts' ratings were either 'Richest', 'Rich' or 'Middle' in the DHS quintiles. On the other hand, a significant proportion of the 'Richest' clusters in the DHS quintiles were rated 'Middle' and 'Rich', with a smaller proportion as 'Poor', with none being rated 'Poorest'. Similarly, none of the clusters in the 'Poorest' quintile of the DHS was rated as 'Richest' by experts, with the largest proportion of the former being rated 'Poor' by the experts. Overall, despite some level of scatter, especially in the middle quintiles, the two independent wealth ratings largely agree on which clusters are the poorest and wealthiest.

To further analyse the relationship between the two sets of median wealth ratings, we run the non-parametric Wilcoxon Signed-Ranks Test. The null hypothesis for this test is that the mean difference between the median expert ratings of cluster images and wealth ratings from the DHS welfare quintiles at the cluster level is significantly different from zero. First, the test revealed that of the 608 pairs of observations, 115 of the clusters had a tied ranking for both quintiles. In other words, 19% of the clusters had the same ratings from the independent DHS and expert ratings quintiles. Second, the experts' wealth ratings were slightly lower (Md = 2.50, n = 358) compared to the DHS wealth ratings (Md = 3.00, n = 135), $z = -11.32$, $p = <0.001$, with a moderate effect size, $r = -0.32$. This means that for 358 of the 608 clusters, human experts estimated slightly higher poverty levels from the imagery compared to the ground truth data based on face-to-face interviews of households within each cluster. This also means that for 135 clusters, DHS ratings for relative poverty levels were lower than expert web ratings based on ocular estimation from imagery. The statistically significant p-value of < 0.001 suggests that the null hypothesis that there is no difference between the two medians is however rejected and that we find a significant difference between the two rankings.

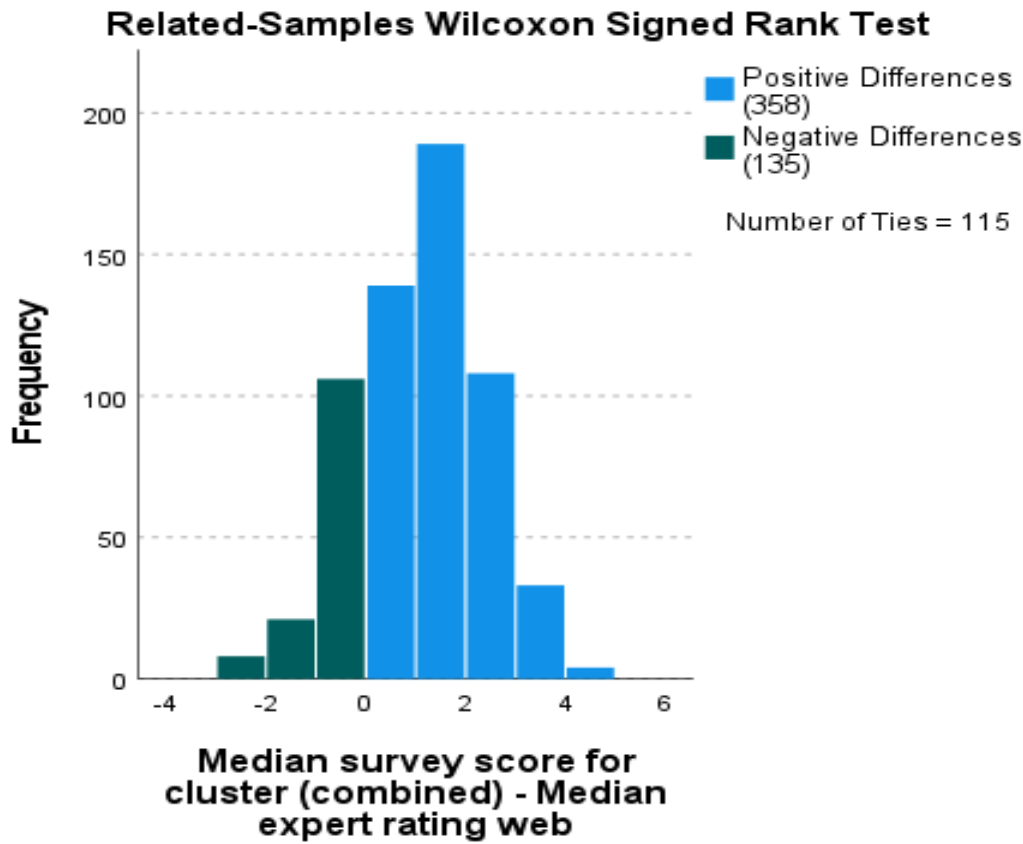

**Figure 2**: A test of the relationship between median experts' and DHS ratings of clusters using the Wilcoxon Signed Rank Test (z = -11.32, p = <0.001).

### 4.3 Covariates of experts' web ratings

Having found a decent relationship between the web ratings of welfare by experts and wealth quintiles from the ground truth data to be statistically significant, and more importantly, the latter to be more realistic of the poverty situation on the ground, we sought to ascertain the key features which can be visually inferred from the images and which were related to the experts' ratings. We employed an ordinal logistic regression model to further understand the relationship between the wealth ratings and the recoded independent variables in Table 2. In the analysis we see that the Pearson chi-square test result of $[X^2(2031) = 1649.509, p = 1.00]$ and the Deviance test $[[X^2(2031) = 1167.660, p = 1.00]$, together with intercept $[X^2(21) = 209.827, p = <0.001]$, suggest the model fit good for the dataset.

Table 5: Regression of experts' welfare ratings on visual features of cluster imagery

| | Variables | Estimate (Std. Error) | p-values | 95% confidence interval | |
|---|---|---|---|---|---|
| | | | | Lower bound | Upper bound |
| Threshold | Poorest quintiles | -3.930 (0.791) | <0.001 | -5.480 | -2.379 |
| | Poor quintiles | -1.065 (0.769) | 0.166 | -2.573 | 0.443 |
| | Middle quintiles | 1.149 (0.772) | 0.137 | -0.364 | 2.662 |
| | Rich quintile | 4.051 (0.833) | <0.001 | 2.419 | 5.683 |
| Building characteristics | Building sizes | 0.275 (0.225) | 0.222 | -0.167 | 0.717 |
| | Modern (slates/tiles) roofs | 0.917 (0.279) | 0.001 | 0.369 | 1.465 |
| | New roofs presence | -0.378 (0.176) | 0.031 | -0.723 | -0.034 |
| | Taller buildings presence | 0.415 (0.222) | 0.062 | -0.021 | 0.850 |
| Landscape features | Low building density | -1.695 (0.374) | <0.001 | -2.427 | -0.963 |
| | Medium building density | -0.565 (0.256) | 0.028 | -1.068 | -0.063 |
| | Low greenery coverage | -0.738 (0.313) | 0.018 | -1.351 | -0.126 |
| | Medium greenery coverage | -0.775 (0.207) | <0.001 | -1.180 | -0.371 |
| Assets and infrastructure | Tarred roads | 0.222 (0.627) | 0.724 | -1.007 | 1.450 |
| | Wider roads | 0.494 (0.218) | 0.023 | 0.068 | 0.921 |
| | Low road coverage | -0.585 (0.214) | 0.006 | -1.004 | -0.165 |
| | Presence of small farms | -0.011 (0.263) | 0.966 | -0.526 | 0.504 |
| | Presence of vehicles | 0.071 (0.196) | 0.717 | -0.313 | 0.455 |

Notes: Only 9 (1.5%) of the clusters were rated 'Richest' so the model removed this threshold. Redundant parameters were also set to zero.

As the regression coefficients from Table 5 show, on the one hand, the presence of modern roofs (slate and clay tile roofing) and wider roads are significant positive predictors of higher scores on the wealth ratings by a human expert. We find the log odds of receiving a rating in a higher quintile on the wealth rankings was 0.917 points higher on average for those clusters which have slate and tile roofing present compared to those which did not. Similarly, the log odds of being rated in a higher wealth quintile is 0.494 points higher on the average for clusters which have wider roads which appear motorable all-year-round. In general terms, as the presence modern roofs and wider roads increases, the probability of a cluster being rated richer increases as well. On the other hand, the main predictor variables with negative and significant regression coefficients include poor road coverage, low and medium greenery coverage, and low and medium building density. For these predictors, as they increase, the probability of a cluster being rated higher on the wealth rankings decreases. Predictors which surprisingly proved non-significant include building sizes, spatial arrangement (residential homes being nucleated, scattered or gridded), whether the main road is tarred or untarred, the presence or absence of farms and their sizes, and the presence of vehicles.

To further model the relationship between the specific wealth quintiles – *poorest*, *poor*, *middle*, *rich* and *richest* – and the visual features, we employed a multinomial logistic regression model. The results show some further nuances in the key covariates for the different categories of welfare. To start with, we see that the model containing the full set of predictors represent a significant improvement in fit relative to the null model [LR$\chi^2$(84) = 275.944, *p*=<0.001]. The pseudo-$R^2$ values as well as the non-significance of the Goodness-of-Fit estimates together suggest a well-fitting model with an overall classification accuracy of 51.2%. Based on the Likelihood Ratio Test results, the key predictors for the whole model include the general settlement structure (gridded, clustered and dispersed), building density, level of greenery, sizes and coverage of roads, and farm sizes. Surprisingly, the presence of vehicles, dominant land *use* (agricultural, commercial, bare lands, and the built-up area), building sizes and roofing status though positive, were non-significant predictors.

Table 6: Multinomial regression results showing the key covariates for poor to richest clusters

| *Cluster category* | *Variable* | *B* | *Std. Error* | *Wald* | *sig.* | Exp (B) | (95% c.i.) Lower bound | Upper bound |
|---|---|---|---|---|---|---|---|---|
| **Poor** | Intercept | 3.011 | 1.634 | 3.395 | 0.065 | | | |
| | Low building density (<25% coverage) | -1.994 | 1.033 | 3.73 | 0.053 | 0.136 | 0.018 | 1.03 |
| | Greenery coverage <25% | -1.79 | 0.788 | 5.165 | 0.023 | 0.167 | 0.036 | 0.782 |
| | Greenery coverage 25 - 75% | -1.264 | 0.613 | 4.254 | 0.039 | 0.283 | 0.085 | 0.939 |
| **Middle** | Intercept | 2.508 | 1.913 | 1.718 | 0.19 | | | |
| | Low building density (<25% coverage) | -3.514 | 1.06 | 10.996 | <.001 | 0.03 | 0.004 | 0.238 |
| | Greenery coverage <25% | -1.854 | 0.815 | 5.183 | 0.023 | 0.157 | 0.032 | 0.773 |

|  | | B | S.E. | Wald | Sig. | Exp(B) | 95% c.i. lower | 95% c.i. upper |
|---|---|---|---|---|---|---|---|---|
|  | **Greenery coverage 25 - 75%** | -1.603 | 0.63 | 6.469 | 0.011 | 0.201 | 0.059 | 0.692 |
|  | Presence of large roads | 2.686 | 1.123 | 5.717 | 0.017 | 14.667 | 1.623 | 132.561 |
| **Rich** | Intercept | 4.659 | 1.858 | 6.283 | 0.012 |  |  |  |
|  | Presence of new roofs | 1.069 | 0.358 | 8.897 | 0.003 | 2.912 | 1.443 | 5.876 |
|  | Low building density (<25% coverage) | -3.962 | 1.17 | 11.474 | <.001 | 0.019 | 0.002 | 0.188 |
|  | Greenery coverage <25% | -2.509 | 0.909 | 7.626 | 0.006 | 0.081 | 0.014 | 0.483 |
|  | Greenery coverage 25 - 75% | -2.214 | 0.669 | 10.96 | <.001 | 0.109 | 0.029 | 0.405 |
|  | Presence of large roads | 3.011 | 1.148 | 6.875 | 0.009 | 20.313 | 2.139 | 192.895 |
|  | Low roads coverage | -1.366 | 0.581 | 5.525 | 0.019 | 0.255 | 0.082 | 0.797 |
| **Richest** | Intercept | -12.044 | 3.815 | 9.967 | 0.002 |  |  |  |
|  | Greenery coverage 25 - 75% | -4.282 | 1.167 | 13.459 | <.001 | 0.014 | 0.001 | 0.136 |
|  | Tarred roads | 11.462 | 1.089 | 110.862 | <.001 | 95066.09 | 11255.927 | 802915.722 |
|  | Cox and Snell | 0.367 |  |  |  |  |  |  |
|  | Nagelkerk | 0.398 |  |  |  |  |  |  |
|  | McFadden | 0.178 |  |  |  |  |  |  |

Notes: Reference category: Poorest, df = 1, c.i.: confidence interval

As Table 6 shows, at the individual wealth ranking levels, variables with negative and significant predictive power include greenery coverage, buildings density and roads coverage while positive and significant predictors include size of roads, road surface, and roof surface quality. The likelihood of clusters being categorized as 'Rich' increases as the presence of large roads increases while the chances of clusters being rated as 'Richest' increases with the increasing presence of tarred roads. For instance, the log-odds of being categorised in the 'Rich' group for the presence of large roads are predicted to be 3.011 points greater than in the 'Poorest' group. Similarly, the log-odds of a clusters with tarred roads being categorised in the 'Richest' group is predicted to be 11.462 points greater than the reference category 'Poorest' category. Positive but non-significant predictors of

# 5. Discussions

## 5.1 Data and methodological lessons
The first relates to the sample size and its implications for the kind of analyses that could be done. Our approach limited us to a relatively small pool of domain experts of 102 respondents which yielded 2,174 ratings for 608 cluster images. This relatively small sample size constrained our ability to test other hypotheses. For example, we expect the accuracy of ratings to be affected by, among other things, educational attainment, length and region of experience of the domain experts. *Ceteris paribus*, we would expect that domain experts with higher educational qualifications, longer fieldwork experience, and main area of fieldwork experience being in East Africa in general and Tanzania in particular, would have better accuracy than other domain experts. A larger sample size could have been achieved by including contacts of our networks to participate in a snowballing approach as well as allowing the online survey to run for a longer period. We could also have promoted the survey at conferences and other similar for where development researchers and practitioners congregate.

The second lesson relates to finetuning of ratings through the training of domain experts. Our domain experts did not receive any training on what markers of poverty to look out for in their ratings. This meant that the criteria were left to each domain expert. We expect that each expert uses the first few ratings as 'target practice' to get the hang of the relative welfare levels in the cluster imagery as they are only presented with one imagery at a time. A future study with a larger number of ratings per cluster could discard the first few ratings of each expert as training data as was done by Tschandl et al. (2019) using a similar framework of human readers on images of skin lesion.

## 5.2 Accuracy of domain experts

Despite the aforementioned limitations, our domain experts come with notable depth and experience, perhaps, thanks to our preferred approach of carefully selecting respondents with the requisite qualifications and relevant fieldwork experience. The other options of sampling may not have yielded this rich sample. A cursory comparison of the distribution of the wealth quintiles for both independent measures follow an inverted U-shaped curve, with steeper edges and a higher middle for the curve of the experts' ratings (Figure 2a). This means that domain experts rated the majority of the clusters in the middle quintiles rather than as richest or poorest. To give some context, despite significant progress in poverty reduction in Tanzania in the last decade, poverty levels, particularly in terms of absolute numbers, are still relatively high. For example, while the national poverty rate has fallen from 34.4% in 2007 to 26.4% in 2018 and extreme poverty rate from 12% to 8% within the same period, as much as 26 million people - 49% of the population – still live below the $1.90 per person per day threshold (Belghith et al., 2019; WorldBank, 2022). The country is thus one of the few countries that the absolute numbers of the poor continue to rise chiefly because population is outstripping economic growth. This background ties in well with the distribution of the wealth quintiles based on ratings by the domain experts.

Interestingly, it appears the wealth quintiles based on the DHS attributes lower levels of poverty than domain experts. For instance, while all the clusters rated 'Richest' by domain experts were also rated 'Richest' by the DHS ratings, the reverse is not the case. Most of the clusters rated as 'Richest' in the DHS were classified as 'Rich' or 'Middle' by domain experts. Conversely, most of the clusters rated 'Poorest' by domain experts were rated 'Poor' or 'Middle' bin the DHS wealth quintiles. This means that estimating welfare levels based on the DHS ratings alone would paint a picture of a more affluent nation. Our finding of relatively lower welfare ratings by human experts is not too surprising given that domain experts rated images based on what is discernible from visually examining the photos. These images are likely to only show assets, social services and infrastructure at the neighbourhood level. These would include roads and roofs, and the quality of these features. The domain experts, thus, did not have the benefit of looking under roofs to count the number of consumer goods such as radios, refrigerators, televisions, and mobile phones which relatively well-off households tend to spend their additional on (Östberg et al., 2018). These are the household assets that DHS enumerators count and assess and which feed into the wealth ranking. Other assets such as vehicles may, with closer scrutiny, be visually discernible from the images presented to the domain experts but surprisingly, this did not show from our regression model as significant. On the one hand, the presence of more modern roofs – houses roofed using slate and tiles – and wider roads are the main covariates of positive and significant higher ratings. On the other hand, poor road coverage being the most important covariate of high poverty ratings is most telling. This is in tandem with the local context where limited access to basic services and assets are endured by poor households and neighbourhoods in Tanzania (Belghith et al., 2019).

The ability of the domain experts to more accurately categorize clusters based on their images for the poorest and richest quintiles is testament to the neighbourhood effect which culminates in socio-spatial fragmentation and segregation whereby the richest and poorest tend to self-isolate (Liu et al., 2017; Otero et al., 2021; van Ham et al., 2012). In Tanzania, like in most other countries in the Global South, this often leads to the sprawling of slums for the ultra-poor and the emergence of well-planned and serviced areas for the rich (Rath, 2022). Among other characteristics, slums often have higher population densities (which could show in images as clustered settlements), low levels of greenery, poor roads quantity and coverage, and disproportionately high dead-end streets (Mahabir et al., 2020). They are also often characterised by temporary housing structures which may show up in images with poor roofing materials. This is in contrast to wealthier neighbourhoods where one would expect to see the opposite – socially and infrastructurally well-serviced homes with better and more durable roofing systems and generally well-planned neighbourhoods. Our finding that clusters which have slate and tile roofs and wider roads tended to receive higher ratings by domain experts is in tandem with those of a longitudinal study by Östberg et al. (2018) of two Tanzanian villages in which they found that residents often invested increased incomes in improved housing and that studying apparently mundane features such as materials used in roofing and farms sizes helps in understanding welfare dynamics at the village level.

## 5.3 Implications for better training of machine learning models

While our correlation coefficients are not particularly impressive, especially when compared to ML approaches which have achieved accuracies as high as 80-90% (See for example Burke et al., 2021; Lee & Braithwaite, 2022), our results can help broaden the frontiers of the use of machine learning and satellite imagery to estimate poverty and welfare by helping to explain the basis for such results. Despite such impressive results, a major shortfall of such studies has been their largely opaque nature and lack of attempts at explainability and reproducibility. This is where our approach can significantly contribute to explaining the inherently black boxes that ML models have tended to be. Similar to our domain experts, ML models do not see below the roofs of buildings. Having recently found, from a review of studies in this domain (Hall et al., forthcoming), that ML models' performance do not diminish with different levels of details, gaining a sneak peek at what these models 'see' and use to achieve such impressive results will contribute to transparency. Unlike domain experts who receive no training on the specific task at hand, ML models get to train on a part of the dataset – the training dataset – with specific target in mind. In some sense, one can view domain experts as possessing general knowledge of their domains while ML models a trained to gain specific knowledge and are then further fine-tuned to achieve the best possible result. Complementing these two systems to estimate welfare levels will maintain the high performance already attainable with ML alone but with better explainability. This could be achieved by training ML models using the key covariates found here to be important for wealth ratings. Apart from roofing quality and roads width and quality, the presence of vehicles has been shown to be important for welfare scoring for ML models (Ayush et al., 2020). We hypothesize that combining these two approaches will augments poverty predictions and lead to explainable AI and eventually causality.

## 6. Conclusions

In the present study, we sought to test domain experts' ability to gauge poverty rates at the neighbourhood level through the visual inspection of satellite imagery. We also sought to shed light on the key features that determine how these images are rated on a five-point poverty scale – 'Poorest', 'Poor', 'Middle', 'Rich', and 'Richest' – in line with the wealth quintiles used by the DHS. To achieve this we set up an experimental survey system in which satellite imagery covering 608 DHS cluster locations in Tanzania were loaded and development researchers with myriad experience and expertise were invited to rate the relative poverty level they could infer by visually examining the images.

We show that domain experts' poverty ratings, by merely examining satellite images of the clusters, are similar to the real poverty situation on the ground in Tanzania. The curve follows an inverted U-shape, with most of the clusters falling within the middle quintiles of 'Poor' and 'Middle', and the least number of clusters being rated as 'Richest'. Tellingly, all the clusters rated as 'Richest' by the domain experts were also rated as 'Richest' in the DHS wealth ranking quintiles. Similarly noteworthy is the fact that none of the 'Poorest' clusters, according to the independent ratings by our domain experts could be found in the 'Richest' quintile of the DHS dataset. This suggests that despite the fact that domain experts do not see below the roof where most household assets can be found, they are, by and large, able to accurately estimate poverty levels merely by examining neighbourhood characteristics. This shows that neighbourhood features such as the quality of roofing materials, roads coverage and width and housing density are key determinants of the level of poverty in a cluster. Our finding that neighbourhoods where tiled and/or slate roofing materials predominate and clusters which have wider road networks tended to be relative well-off is instructive. Equality important is our finding that clusters in which key features of slums – high density of temporary looking structures with poor road coverage – tended to receive higher poverty ratings. This knowledge holds great potential for more ML models which till date achieve impressive results in poverty analysis but with little transparency as to what features are used by the models. Going forward, features found to be strongly associated with poverty ratings could be useful for training deep and transfer learning models and thus overcome a major challenge of training models – the lack of adequate labelled data. By so doing, we contribute to improving the explainability of such predominantly inexplicable but impressive results at the intersection of satellite imagery, artificial intelligence and poverty analysis.

# Acknowledgement

The authors would like to thank financial support during the project from the Swedish Research Council 2019-04253, Riksbankens Jubileumsfond MXM19-1104:1 and the Crafoord foundation.